\documentclass[aps,prl,twocolumn,longbibliography,superscriptaddress]{revtex4-2}
\usepackage{bbm}
\usepackage{graphicx}
\usepackage{dcolumn}
\usepackage{bm}
\usepackage{subfigure}
\usepackage{amsmath}
\usepackage{feynmf}
\usepackage{hyperref}
\usepackage{CJK}
\usepackage{amssymb}
\usepackage{attachfile}
\newcommand{\bk}{\boldsymbol k}
\newcommand{\bp}{\boldsymbol p}

\usepackage{braket}
\usepackage{esint}
\usepackage{times}

\begin{document}

\title{Nonrelativistic Spin-Orbit-Coupling Effects in Odd-Parity Coplanar Magnets}

\author{Dongling Liu}
\affiliation{Guangdong Provincial Key Laboratory of Magnetoelectric Physics and Devices,
State Key Laboratory of Optoelectronic Materials and Technologies,
School of Physics, Sun Yat-sen University, Guangzhou 510275, China}

\author{Zheng-Yang Zhuang}
\affiliation{Guangdong Provincial Key Laboratory of Magnetoelectric Physics and Devices,
State Key Laboratory of Optoelectronic Materials and Technologies,
School of Physics, Sun Yat-sen University, Guangzhou 510275, China}

\author{Di Zhu}
\affiliation{Guangdong Provincial Key Laboratory of Magnetoelectric Physics and Devices,
State Key Laboratory of Optoelectronic Materials and Technologies,
School of Physics, Sun Yat-sen University, Guangzhou 510275, China}

\author{Zhigang Wu}
\affiliation{Quantum Science Center of Guangdong-Hong Kong-Macao Greater Bay Area (Guangdong), Shenzhen 508045, China}

\author{Zhongbo Yan}
\email{yanzhb5@mail.sysu.edu.cn}
\affiliation{Guangdong Provincial Key Laboratory of Magnetoelectric Physics and Devices,
State Key Laboratory of Optoelectronic Materials and Technologies,
School of Physics, Sun Yat-sen University, Guangzhou 510275, China}

\date{\today}

\begin{abstract}
Spin-orbit coupling (SOC) is a relativistic effect that underpins a broad spectrum of phenomena 
in condensed matter physics, from topological phases of matter to spintronic functionality. Its 
relativistic origin, however, restricts strong SOC to heavy-element materials 
and locks spin-momentum texture into a fixed, material-specific pattern.
Here we show that odd-parity coplanar magnets offer a nonrelativistic pathway to highly tunable SOC effects. 
We construct a bilayer coplanar magnet via symmetry-guided stacking of two monolayer odd-parity altermagnets and demonstrate that Rashba, 
Weyl, and Dresselhaus spin textures can all be realized, 
and that the spin texture can be switched between these forms simply by tuning the layer N\'{e}el vector.
Through the spin Edelstein effect and the realization of fully gapped chiral topological superconducting phases, we demonstrate that 
this nonrelativistic SOC achieves physical equivalence to its relativistic counterpart.
Our findings identify a new class of odd-parity coplanar magnets as a versatile platform for engineering SOC effects. 
\end{abstract}

\maketitle
{\it Introduction.---}Physics arising from spin-orbit coupling (SOC) has been a central theme 
in condensed matter physics over the past decades~\cite{roland2003spin,Galitski2013,Manchon2015}, giving rise to a wide range of phenomena 
such as various Hall effects~\cite{Nagaosa2010RMP,Sinova2015,Chang2023QAHE}, unconventional superconductivity~\cite{bauer2012non,Smidman2017}, and topological phases of matter~\cite{hasan2010,qi2011,Armitage2018RMP}. 
This rich physics stems from two key effects of SOC on Bloch bands. First, SOC leads 
to spin-momentum locking, which generates nontrivial quantum geometry in the wave 
functions~\cite{Giacomo2025} and fundamentally affects electron transport behaviors~\cite{Xiao2010review,Bercioux2015}. 
Second, in systems lacking inversion symmetry, SOC induces odd-parity momentum-dependent 
spin splitting, resulting in spin-split Fermi 
surfaces with intriguing spin-texture patterns~\cite{Zhang2025SML}. Rashba, Weyl (also called radial Rashba), Dresselhaus 
SOC are three basic types that emerge in systems globally or locally lacking inversion symmetry~\cite{Zhang2014}, 
each having generated extensive research due to its distinct spin-momentum locking~\cite{Ganichev2004,Bernevig2006helix,Koralek2009,Manchon2009,Caviglia2010,Lin2011SOC,
Wang2012SOC,Niesner2016SOC,Rowland2016,Bihlmayer2022,Gatti2020,Kang2024radial,Costa2025,Sousa2025}. 
Because they respect different symmetries, these three types of SOC typically arise in distinct 
classes of materials and switching between them in a given material is 
not feasible.

Although SOC is ubiquitous in materials, its relativistic origin confines 
strong SOC primarily to heavy-element systems. This limitation has motivated 
the search for a nonrelativistic analog that realizes the equivalent effects of SOC.
To date, two main nonrelativistic mechanisms have been established. The first is an 
interaction-driven spin-channel Pomeranchuk instability of the Fermi surface~\cite{Wu2004SOC,We2007FL}. 
In this mechanism, SOC emerges as a spin–orbit order parameter. The second mechanism arises 
from the exchange interaction between itinerant electrons and a noncollinear magnetic order background~\cite{Fujita2011,Martin2012HM}. 
Through a local gauge transformation that aligns all on-site exchange fields along a given direction, the 
spatial variation of the exchange field orientation can be captured as an emergent SU(2) non-Abelian gauge field felt by 
the itinerant electrons~\cite{Fujita2011}, which consequently leads to effects equivalent to those of SOC.
While the first mechanism awaits experimental realization in materials, the second has already found material realizations~\cite{Masaya2006,Togawa2012,Togawa2016,Song2025pwave,Yamada2025}. 

The recently discovered odd-parity magnets provide an explicit realization of the second mechanism.
These magnets originally refer to coplanar antiferromagnets with an effective time-reversal 
symmetry $\mathcal{T}\boldsymbol{\tau}$ ($\mathcal{T}$ is time-reversal, and $\boldsymbol{\tau}$ denotes 
a fractional translation)~\cite{Birk2023OPM}, 
which enforces zero net magnetization and odd-parity spin polarization.
This has triggered extensive research~\cite{Brekke2024pwave,
Maeda2024,Ezawa2025pwave,Ezawa2025Edelstein,Sukhachov2025,Yu2025OPM,Sun2025pwave,Zeng2025pwave,Tim2025pwave,
Chakraborty2025,Lee2026OPM,Fu2026Floquet,Zhang2026pwave,Dsouza2026,
Priessnitz2026,Carmelo2026OPM,Cuono2026pwave,Yuan2026pwave,Mitscherling2026pwave,Li2026pwave,Sim2026pwave,
Fakhredine2026pwave,Froldi2026pwave,Leeb2026pwave,Hirschmann2026OPM,Fukaya2025pwave,Salehi2025pwave,Luo2025SGS,Luo2026SGS}.  
However, the effective SOC in these materials is one
dimensional as the spin polarization is unidirectional: for magnetic moments in the $xy$ plane, 
symmetry constrains $\langle s_{x,y}(\bk)\rangle = 0$ and $\langle s_{z}(\bk)\rangle = -\langle s_{z}(-\bk)\rangle\neq0$
at  generic momenta~\cite{Song2025pwave,Yamada2025}.  This restriction precludes many SOC-driven 
effects and topological phases in dimensions greater than one. For example, in two dimensions, combining this effective 1D SOC 
with spin-singlet pairing cannot yield a fully gapped topological 
superconducting phase, but only topological nodal superconducting phases~\cite{Nagae2025OPM,Amartya2026,Kim2026OPM,Luo2026OPM}. 
This raises a fundamental question: can higher-dimensional nonrelativistic SOC effects be realized within an odd-parity coplanar magnet?

In this Letter, we answer affirmatively  by identifying a new class of  odd-parity coplanar magnets
characterized by  $\langle s_{x,y}(\bk)\rangle = -\langle s_{x,y}(-\bk)\rangle\neq0$ and $\langle s_{z}(\bk)\rangle = 0$. 
We establish the symmetry conditions for their existence and design a bilayer framework for their realization. 
Remarkably,  we find that these two odd-parity spin polarization components can form topologically winding spin textures identical to those induced 
by Rashba, Dresselhaus, and Weyl SOC. Moreover, these distinct spin textures can be  switched into one another 
by tuning the N\'{e}el vector.
Through the spin Edelstein effect and the realization of fully gapped chiral topological superconducting phases, 
we demonstrate that this nonrelativistic 2D SOC achieves physical equivalence to its relativistic counterpart 
while offering a degree of tunability that the latter fundamentally lacks.

{\it Symmetry analysis and general framework.---}We first establish the symmetry conditions for $\langle s_{x,y}(\bk)\rangle = -\langle s_{x,y}(-\bk)\rangle\neq0$ and $\langle s_{z}(\bk)\rangle=0$ at generic momenta in a coplanar antiferromagnet with $xy$-plane magnetic moments. 
The first condition 
is the existence of symmetries that enforce the spin polarizations to be in the $xy$-plane and odd-parity. 
The required symmetries can be $[C_{2\perp}\Vert\mathcal{P}]$ (for 2D systems, $[C_{2\perp}\Vert C_{2z}]$ also 
suffices) and an effective time-reversal symmetry 
$\mathcal{T}\boldsymbol{\tau}$.
Here we adopt the usual spin group notation where the operations to the left and  right of $\Vert$ act on the spin and spatial degrees of freedom, respectively. For example, $[C_{2\perp}\Vert\mathcal{P}]$ means the symmetry under a combined operation of a $\pi$ spin rotation about an axis perpendicular to the spin and a spatial inversion. When the aforementioned two symmetries coexist,  they impose opposite constraints on the parity of $\langle s_z(\bk)\rangle$ so that $\langle s_z(\bk)\rangle= 0$. The effective time-reversal symmetry also ensures the odd-parity of the $xy$-plane spin polarization, i.e., $\langle s_{x,y}(\bk)\rangle = -\langle s_{x,y}(-\bk)\rangle$. 
The second condition is the absence of all symmetries that enforce even-parity in-plane spin polarizations, specifically the inversion symmetry $\mathcal{P}$ and the symmetry $[\bar{C}_{2\perp}\Vert\mathcal{T}]$ (where the overbar indicates an additional spin reversal induced by time reversal).  This ensures that $\langle s_{x,y}(\bk)\rangle \neq 0$. Notably, $[\bar{C}_{2\perp}\Vert\mathcal{T}]$ 
is always present in coplanar antiferromagnets if the bare Hamiltonian (without spin magnetic order) 
is time-reversal symmetric~\cite{Zhu2026Even}. This symmetry can be broken by applying circularly polarized light~\cite{huang2025oddparityAM,li2025floquet,zhu2025floquet,
liu2025floquet,Zhu2026Even,Zhuang2026mixed} or if the system hosts 
charge loop currents~\cite{lin2025,Zeng2025OPAM,zeng2025} or complex orbital orders~\cite{zhuang2025odd}.
These four symmetries---the two that must be absent and the two that must be present---constitute the complete minimal set required.

\begin{figure*}[t]
	\centering
	\includegraphics[width=0.8\textwidth]{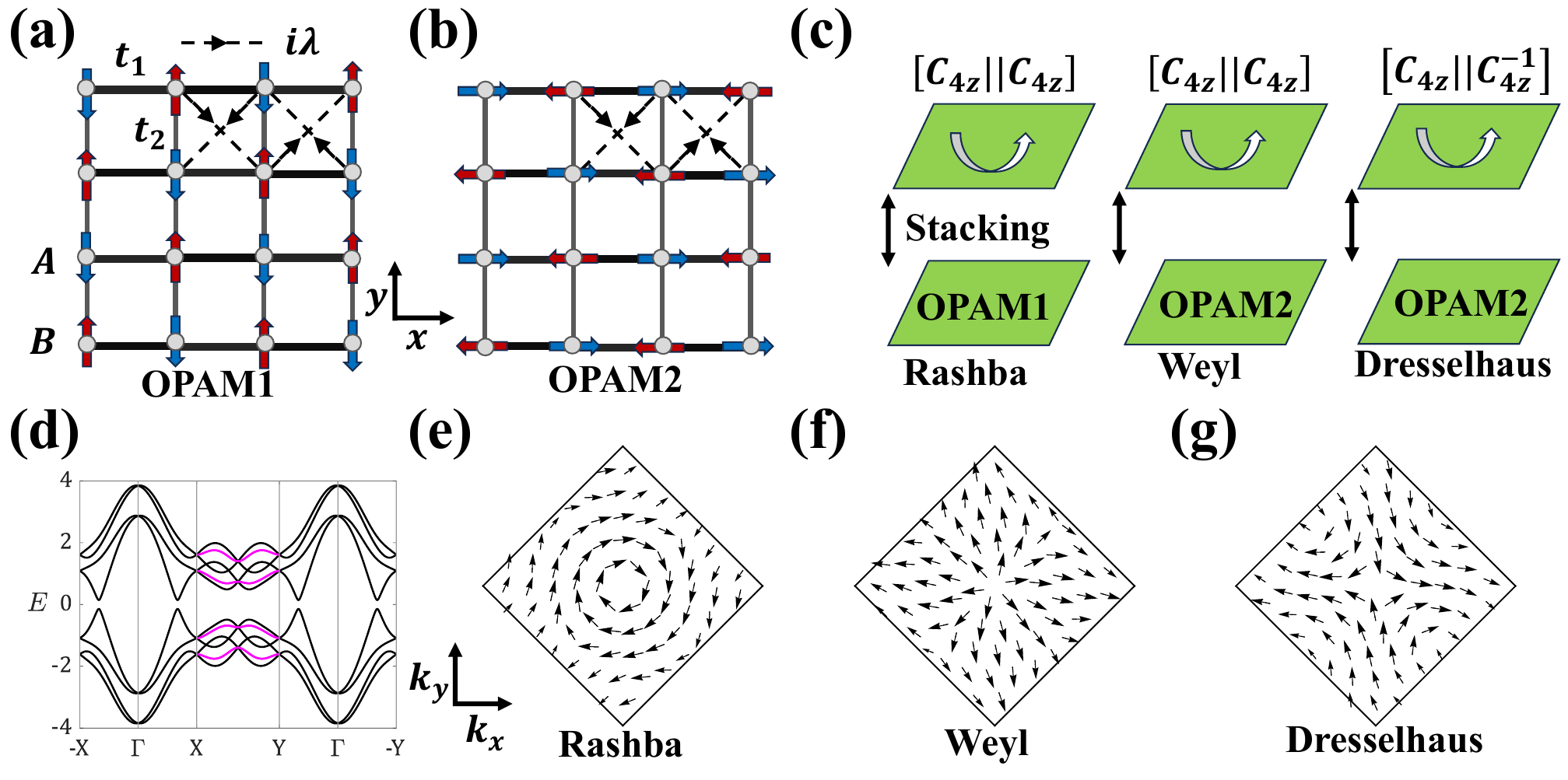}
	\caption{(a) and (b) Two spin configurations for the bottom layer, OPAM1 and OPAM2, differing by a $90^{\circ}$ of the N\'{e}eel vector. Red/blue arrows: opposite magnetic moments on sublattices A/B. Solid (dashed) lines: nearest-neighbor (next-nearest-neighbor) hoppings. 
     (c) Symmetry-guided stacking for realizing effective Rashba (left),  Weyl (middle), and  Dresselhaus  (right) SOC.
     The top layer is obtained from the bottom layer by the symmetry operation indicated in brackets.
     (d) Energy spectra: black lines (Rashba and Weyl cases, fully overlapped) and magenta lines (Dresselhaus case). The spectra for all three cases are identical along the -X-$\Gamma$-X and -Y-$\Gamma$-Y directions.
     (e–g) Spin textures of the lowest-energy band for the three distinct stacking configurations outlined in (c). Parameter conditions 
     are identical, with $t_{1}=1$, $t_{2}=0.6$, $\lambda=0.25$, $M=1$, and $\eta=0.5$.
      }\label{Fig1}
\end{figure*}

To fulfill these symmetry conditions, we propose the bilayer stacking strategy sketched in Figs.~\ref{Fig1}(a--c). 
Specifically, the bottom layer is a square-lattice odd-parity altermagnet (OPAMs)~\cite{liu2025floquet} lacking $\mathcal{P}$ and $[\bar{C}_{2\perp}\Vert\mathcal{T}]$ but preserving
$[C_{2\perp}\Vert C_{2z}]$ and $\mathcal{T}\boldsymbol{\tau}$; its spin polarization 
is odd-parity and unidirectional along the N\'{e}el vector.
The top layer is related to the bottom layer by a $90^{\circ}$ spin-lattice locked rotation 
($[C_{4z}\Vert C_{4z}]$ or $[C_{4z}\Vert C_{4z}^{-1}]$ in spin group notation),
rendering its odd-parity spin polarization perpendicular to that of the bottom layer. 
The orthogonal N\'{e}el vectors of the two layers form a coplanar spin configuration.
Upon interlayer coupling, as we demonstrate below, diverse 2D
topologically winding spin textures emerge, with their explicit form determined by the specific 
symmetry relationship between the layers.

{\it Rashba, Weyl and Dresselhaus spin textures.---}The Hamiltonian describing the bilayer magnet takes the general form
\begin{equation}
	\mathcal{H}(\bk)=\begin{pmatrix}
		\mathcal{H}^t(\bk) & \eta\\ \eta & \mathcal{H}^b(\bk)
	\end{pmatrix},\label{HRt}
\end{equation}
where $\mathcal{H}^{t/b}$ describes the top/bottom layer in the decoupled limit, 
and $\eta$ characterizes the interlayer coupling strength for AA stacking. The explicit form of 
the bottom-layer Hamiltonian is
\begin{eqnarray}
\mathcal{H}^b(\bk)&=&2(t_1\cos k_x+t_2\cos k_y)\sigma_x+4\lambda\sin k_x\cos k_y\sigma_z\nonumber\\
		&&+\boldsymbol{M}\cdot \boldsymbol{s}\sigma_z,\label{bottomH}
\end{eqnarray}
where the Pauli matrices $\sigma_{x,z}$ act on the two magnetic sublattice degrees of freedom, and 
the Pauli-matrix vector $s=(s_{x},s_{y},s_{z})$ acts on the spin space. Identity matrices are left implicit and 
lattice constants are set to unity throughout. The first term describes nearest-neighbor hopping between magnetic sublattices, 
the second captures a sublattice current order, and the last term represents the 
exchange coupling with an in-plane exchange field $\boldsymbol{M}=(M_{x},M_{y},0)$. 
The sublattice current breaks $[\bar{C}_{2z}\Vert \mathcal{T}]$; such a term can also 
be induced by circularly polarized light if the lattice is dimerized~\cite{liu2025floquet}.
The exchange coupling breaks inversion symmetry ($\mathcal{P}=\sigma_{x}$) and time-reversal symmetry 
$\mathcal{T}=is_{y}\mathcal{K}$ individually, but preserves the combined $\mathcal{P}\mathcal{T}$ symmetry. 
The full Hamiltonian $\mathcal{H}^b(\bk)$, however, lacks $\mathcal{P}\mathcal{T}$ while preserving
 $[C_{2z}\Vert \mathcal{P}]$ and an effective time-reversal symmetry $\mathcal{T}\boldsymbol{\tau}=\sigma_{x}s_{y}\mathcal{K}$. 
Consequently, it describes an OPAM whose band structure exhibits a unidirectional odd-parity spin polarization 
along the N\'{e}el vector orientation.~\cite{liu2025floquet}.

 We consider two spin configurations 
for the bottom layer, OPAM1 and OPAM2, as sketched in Figs.~\ref{Fig1}(a) and \ref{Fig1}(b). When the 
top layer is related to the bottom layer by $g_{1}=[C_{4z}\Vert C_{4z}]$, i.e., 
$\mathcal{H}^t(\bk)=g_{1}\mathcal{H}^b(\bk)$ or more explicitly $\mathcal{H}^t(k_{x},k_{y})
=e^{i\frac{\pi}{4}s_{z}}\mathcal{H}^b(-k_{y},k_{x})e^{-i\frac{\pi}{4}s_{z}}$,
we find (i) Rashba spin texture for OPAM1 [left panel of Fig.~\ref{Fig1}(c)] and (ii) Weyl spin 
texture for OPAM2 [middle panel of Fig.~\ref{Fig1}(c)]. When instead the layers are related by
$g_{2}=[C_{4z}\Vert C_{4z}^{-1}]$, i.e., $\mathcal{H}^t(\bk)=g_{2}\mathcal{H}^b(\bk)$ with $\mathcal{H}^t(k_{x},k_{y})=e^{i\frac{\pi}{4}s_{z}}\mathcal{H}^b(k_{y},-k_{x})e^{-i\frac{\pi}{4}s_{z}}$,
we find (iii) Dresselhaus spin texture for OPAM2 [right panel of Fig.~\ref{Fig1}(c)]. In practice,
$g_{1}$ corresponds to an  anticlockwise $90^{\circ}$ rotation of the layer as a whole, 
while $g_{2}$ is realized by a clockwise $90^{\circ}$ rotation of the layer combined with a 
reversal of the N\'{e}el vector ($[C_{4z}\Vert C_{4z}^{-1}]=
[C_{2z}\circ C_{4z}^{-1}\Vert C_{4z}^{-1}$]). 

Figure \ref{Fig1}(d) shows the energy spectra along selected high-symmetry lines for 
these three cases. The Rashba and Weyl spectra are identical, while the Dresselhaus spectrum differs by a $90^{\circ}$
rotation in momentum space---a direct consequence of the symmetry relations between their Hamiltonians. 
Figures \ref{Fig1}(e)--\ref{Fig1}(g) present the spin textures of the lowest-energy band for the three cases, 
exhibiting the characteristic Rashba, Weyl, and Dresselhaus patterns, respectively.

To quantify the effective SOC strength, we combine a $\bk\cdot\bp$ expansion with a
two-step projection to derive a low-energy continuum Hamiltonian  that effectively 
describes the lowest-energy pair of bands near the $\boldsymbol{\Gamma}=(0,0)$ point. 
This procedure also yields an analytical expression for the spin texture of the 
lowest-energy band.

We focus on the Rashba case for illustration (Weyl and Dresselhaus cases are provided in the Supplemental Material (SM)~\cite{supplemental}). 
In the decoupled limit ($\eta=0$),
the Hamiltonian exhibits a fourfold degeneracy at $E=-I_{0}$, where $I_{0}=\sqrt{4(t_1+t_2)^2+M^2}$. The four 
corresponding eigenstates can be chosen as  
\begin{eqnarray}
|u_{\alpha}^{a}\rangle=\frac{1}{\sqrt{2}}(|u^{a}_{1}\rangle+\alpha |u_{2}^{a}\rangle),
\end{eqnarray}
where $\alpha=\pm$, $a=\{t,b\}$,  and
$|u^{t(b)}_{j}\rangle=|t(b)\rangle\otimes|s_{x(y)}=(-1)^{j+1}\rangle\otimes|\chi_{j}\rangle$
with $j=1,2$. Here,  $|t(b)\rangle$ denotes full polarization to the top (bottom) layer, 
$|s_{x(y)}=\pm1\rangle$ are the two eigenstates of the Pauli matrix $s_{x(y)}$, and 
$|\chi_{j}\rangle$ is given by
\begin{eqnarray}
|\chi_{j}\rangle&=&\frac{1}{\sqrt{2I_{0}(I_{0}+(-1)^{j}M)}}\left(
                                                \begin{array}{c}
                                                  2(t_{1}+t_{2}) \\
                                                  -(I_{0}+(-1)^{j}M) \\
                                                \end{array}
                                              \right).
\end{eqnarray}
Projecting the full Hamiltonian onto the basis spanned by $(|u_{+}^{t}\rangle,|u_{-}^{t}\rangle,|u_{+}^{b}\rangle,|u_{-}^{b}\rangle)$
and retaining terms up to linear order in momentum and $M/I_{0}$ yields a four-band Hamiltonian:
\begin{eqnarray}
\mathcal{H}_{\rm FB}(\bk)&=&-I_{0}-2\alpha_{\rm so} k_{y}\frac{\rho_{z}+1}{2}s_{x}+2\alpha_{\rm so} k_{x}\frac{1-\rho_{z}}{2}s_{x}\nonumber\\
&&+\frac{\eta}{2}(\rho_{x}-\rho_{y})+\frac{\eta}{2}(\rho_{x}+\rho_{y})s_{z},\label{effectiveRH}
\end{eqnarray}
where $\alpha_{\rm so}=2\lambda M/I_{0}$, and $\rho_{i}$ are Pauli matrices 
acting on a layer-sublattice mixing subspace. At $\boldsymbol{\Gamma}$,  $\mathcal{H}_{\rm FB}$ 
possesses two pairs of degenerate eigenvalues: one at 
$E_{1}=-I_{0}-\eta$  and the other at $E_{2}=-I_{0}+\eta$. The two eigenstates corresponding to $E_{1}$ are 
\begin{eqnarray}
|v_{1}\rangle&=&|\rho_{x}=-1\rangle\otimes|s_{z}=1\rangle, \nonumber\\
|v_{2}\rangle&=&|\rho_{y}=1\rangle\otimes|s_{z}=-1\rangle.
\end{eqnarray}
A second projection onto the basis 
$(|v_{1}\rangle,|v_{2}\rangle)$ yields the low-energy effective Hamiltonian for the lowest-energy pair of bands:
\begin{eqnarray}
\mathcal{H}_{\rm LEP}(\bk)=-I_{0}-\eta+\alpha_{\rm so}(k_{x}s_{y}-k_{y}s_{x}). \label{lowenergyRH}
\end{eqnarray}  
The linear momentum term takes a form identical to Rashba SOC, with $\alpha_{\rm so}=2\lambda M/I_{0}$ characterizing the 
SOC strength. To verify that this Hamiltonian indeed realizes Rashba SOC, we calculate the 
spin texture, which is independent of the gauge choice of basis functions (see SM~\cite{supplemental} for details). 

For comparison with the numerical result shown in Fig.~\ref{Fig1}(e), we focus on the lower 
band of $\mathcal{H}_{\rm LEP}(\bk)$. Its eigenstate is 
$|\psi_{-}(\bk)\rangle=(1,-ie^{i\theta_{\bk}})^{T}/\sqrt{2}$, where $\theta_{\bk}=\arg(k_{x}+ik_{y})$. 
Taking into account the basis functions at each projection step, this eigenstate can be expressed in the 
original eight-band basis as
\begin{eqnarray}
|\psi_{-}(\bk)\rangle=\frac{1}{2}[(|u_{+}^{t}\rangle-|u_{+}^{b}\rangle)-ie^{i\theta_{\bk}}(|u_{-}^{t}\rangle+i|u_{-}^{b}\rangle)].
\end{eqnarray}
The spin polarization $s_{i}(\bk)\equiv\langle\psi_{-}(\bk)|s_{i}|\psi_{-}(\bk)\rangle$ is then~\cite{supplemental} 
\begin{eqnarray}
\langle s_{x}(\bk)\rangle=\sin\theta_{\bk},
\langle s_{y}(\bk)\rangle=-\cos\theta_{\bk},
\langle s_{z}(\bk)\rangle=0.
\end{eqnarray}
The in-plane spin texture $(\langle s_{x}(\bk)\rangle, \langle s_{y}(\bk)\rangle) = (\sin\theta_{\bk}, -\cos\theta_{\bk})$ 
is everywhere perpendicular to the momentum $\bk = k(\cos\theta_{\bk}, \sin\theta_{\bk})$. This is precisely the hallmark of a 
Rashba spin texture.

 {\it Spin Edelstein effect.---}Under an electric field, SOC can induce spin accumulation---a phenomenon 
 known as spin Edelstein effect (SEE)~\cite{DYAKONOV1971,1989Aronov,1990Edelstein}. 
  This charge-to-spin conversion effect plays a fundamental role in spintronics.  Within linear response theory~\cite{Kubo:1957mj}, 
  the spin accumulation decomposes into time-reversal-even and -odd contributions: $\delta \boldsymbol{S}=\delta \boldsymbol{S}^{\text{even}}+\delta \boldsymbol{S}^{\text{odd}}$~\cite{johanssonTheorySpinOrbital2024}. Although our bilayer models
    lack time-reversal symmetry, we find $\delta \boldsymbol{S}^{\text{odd}}$ still vanishes due to the effective time-reversal symmetry $\mathcal{T}\tau$. Therefore,  only $\delta \boldsymbol{S}^{\text{even}}$ remains. Accordingly, we have~\cite{Garate2009,Chakraborty2025}
\begin{equation}
	\delta\boldsymbol{S}=-\frac{e}{V}\sum_{n,\boldsymbol{k}}\delta(\epsilon_{n\boldsymbol{k}}-\epsilon_F)
\tau_{n\boldsymbol{k}}\boldsymbol{s}_{n\boldsymbol{k}}\left(\boldsymbol{v}_{n\boldsymbol{k}}\cdot \boldsymbol{E}\right). \label{SEE}
\end{equation}
Here, $e=-|e|$ is the electron charge, $V$ the system volume, $\boldsymbol{E}=(E_{x},E_{y})$ the applied electric field,  
$n$ the band index, $\epsilon_F$  the Fermi energy, and $\tau_{n\boldsymbol{k}}$ 
the quasiparticle lifetime. The quantities $\boldsymbol{s}_{n\boldsymbol{k}}$ and $\boldsymbol{v}_{n\boldsymbol{k}}$ 
denote the spin polarization and group velocity of the $n$th band at momentum $\bk$. In component form, 
the spin accumulation can be expressed as $\delta S_{i}=\chi_{ij}E_{j}$, where 
$\chi_{ij}$ is the SEE susceptibility tensor.

For the textbook Rashba Hamiltonian $\mathcal{H}_{R}^0(\bk)=\frac{\hbar^2k^2}{2m}+\alpha_{R}(k_xs_y-k_ys_x)$, 
the two mirror symmetries $\mathcal{M}_{x}$ and $\mathcal{M}_{y}$ enforce $\chi_{xx}=\chi_{yy}=0$, 
while the rotation symmetry enforces $\chi_{xy}=-\chi_{yx}$. 
In the regime with $\epsilon_F>0$ (Fermi level above 
the band degeneracy at $\boldsymbol{\Gamma}$),  $\chi_{xy}$ exhibits 
a salient characteristic: under the constant relaxation time approximation ($\tau_{n\bk}=\tau$), 
it becomes $\chi_{xy}=e\tau\alpha_R m/4\pi\hbar^2$~\cite{johanssonTheorySpinOrbital2024}---a constant
linear in  the SOC strength $\alpha_R$. 
Using realistic parameters---$\alpha_R=0.3\, \text{eV}\text{\AA}$ (a typical value of Au(111) surface~\cite{Popovi2005}), 
$\tau=\hbar/2\delta$ with $\delta=0.01\, \text{eV}$, and $m$ taken as the free electron mass $m_e$---we obtain
$\chi_{xy}=-0.17\,\hbar/\text{V\AA}$.

\begin{figure}[t]
	\centering
	\includegraphics[width=0.45\textwidth]{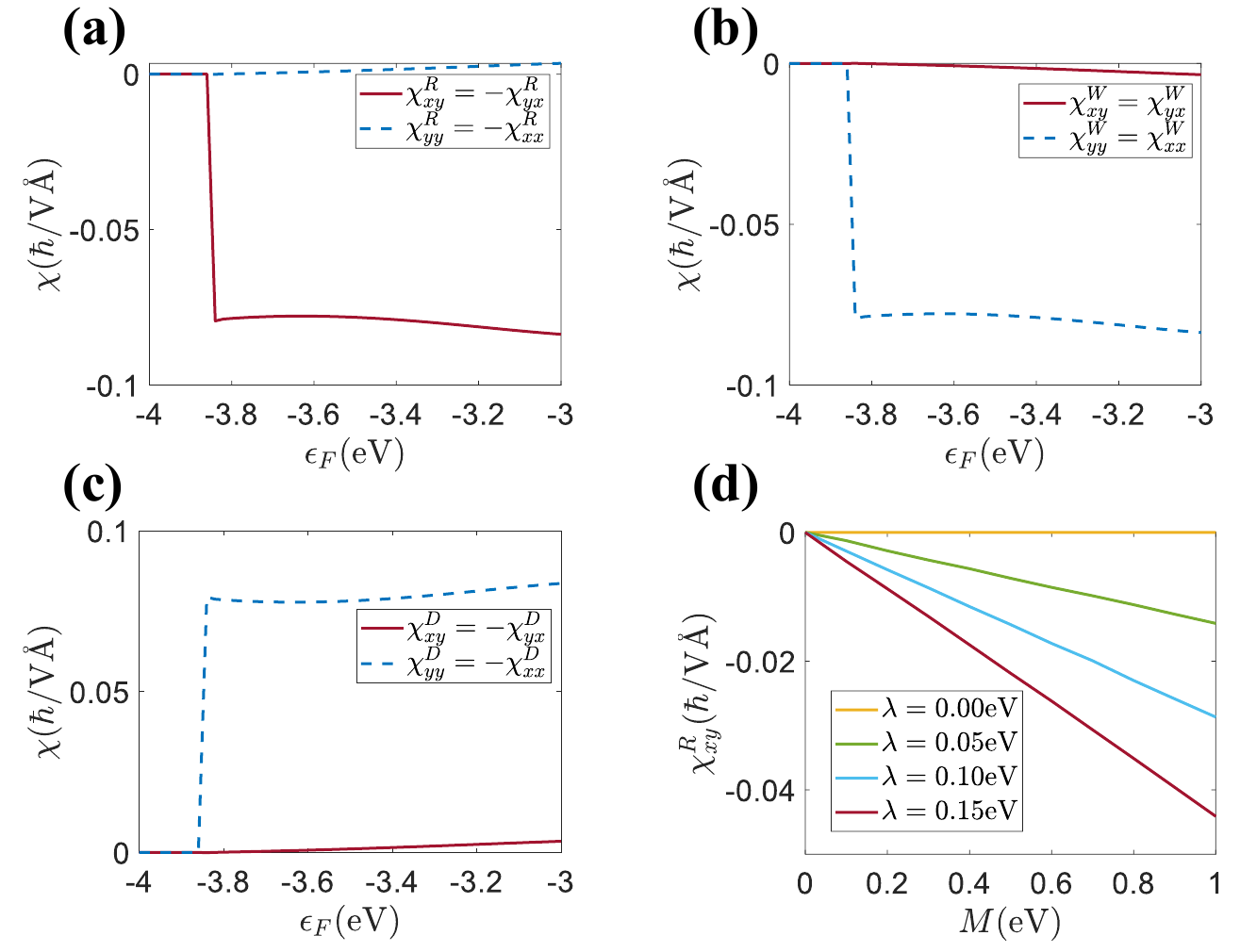}
	\caption{Fermi energy-dependence of the SEE susceptibility tensor components for Rashba (a), Weyl (b), and Dresselhaus (c) case. (d) Dependence of $\chi^{R}_{xy}$ on $M$ for different values of $\lambda$. Shared parameters: $t_1=1.0\, \text{eV}$, $t_2=0.6\, \text{eV}$, $\eta=0.5\, \text{eV}$, $\delta=0.01\, \text{eV}$, lattice constant $a=5\, \text{\AA}$. $\lambda=0.25\, \text{eV}$ and $M=1.0\, \text{eV}$. Under this set of parameters, $\chi_{xy}$ can reach the value $-0.08\,\hbar/\text{V\AA}$.}\label{Fig2}
\end{figure}

Since our bilayer realizes effective Rashba, Weyl, and Dresselhaus SOC, we evaluate 
$\chi_{ij}$ for all three cases under identical conditions. For direct comparison with the textbook model, 
we set the Fermi level to cross the lowest pair of bands. As shown in Fig.~\ref{Fig2}(a) for the Rashba case,
$\chi_{xy}$  rises rapidly and becomes a nearly constant value as the Fermi level moves above 
the band bottom---behavior analogous to the textbook Rashba model. However, 
$\chi_{xy}$ is not perfectly constant when the Fermi level lies above the band degeneracy, and 
$\chi_{yy}$ ($\chi_{xx}=-\chi_{yy}$), despite being very small, 
approaches zero only when the Fermi surface is very close to $\boldsymbol{\Gamma}$. 
These deviations are expected: the lattice Hamiltonian progressively deviates from the 
continuum Rashba form as the Fermi surface expands away from $\boldsymbol{\Gamma}$, and the mirror symmetries 
$\mathcal{M}_{x}$ and $\mathcal{M}_{y}$---absent in the full lattice model---emerge only in the 
low-energy expansion near $\boldsymbol{\Gamma}$. Nevertheless, when the Fermi level 
stays near the band degeneracy at $\boldsymbol{\Gamma}$, the overall behavior remains essentially 
the same as that of the textbook Rashba Hamiltonian.

Figures \ref{Fig2}(b) and \ref{Fig2}(c) present $\chi_{xy}$ and $\chi_{yy}$ for the Weyl and 
Dresselhaus cases. Comparing Fig.~\ref{Fig2}(b) with Fig.~\ref{Fig2}(a), we find the relations: 
$\chi_{xy}^{W}=-\chi_{yy}^{R}$ and $\chi_{yy}^{W}=\chi_{xy}^{R}$
(superscripts: R for Rashba, W for Weyl, D for Dresselhaus). Interpreting $\chi_{xy}$ and $\chi_{yy}$
as two orthogonal coordinates of a 2D vector, this relation indicates that the Weyl susceptibility 
is obtained from the Rashba susceptibility by a $90^{\circ}$ rotation. Comparing Fig.~\ref{Fig2}(c) with Figs.~\ref{Fig2}(a) and \ref{Fig2}(b), we find 
$\chi_{xy}^{D}=-\chi_{xy}^{W}=\chi_{yy}^{R}$ and $\chi_{yy}^{D}=-\chi_{yy}^{W}=-\chi_{xy}^{R}$, meaning that the Dresselhaus vector ($\chi_{xy}^{D},\chi_{yy}^{D}$) 
is related to the Rashba vector ($\chi_{xy}^{R},\chi_{yy}^{R}$)  by an opposite $90^{\circ}$ rotation. 
Figure \ref{Fig2}(d) displays the dependence of 
$\chi_{xy}$ on $M$ for several values of $\lambda$; $\chi_{xy}$ is nearly linear in both $M$ and 
$\lambda$ over the parameter range considered. This agrees with our analytical result $\alpha_{\rm so}=2\lambda M/I_{0}$ 
and the linear scaling $\chi_{xy}\sim\alpha_{\rm so}$  derived for the textbook Rashba model.

{\it Chiral topological superconductors.---}SOC also plays a key role in realizing 
topological superconducting phases~\cite{fu2008,Sato2009TSC,Sau:2009cqc,Lutchyn:2010xey,Oreg:2010xfn,Wong2012majorana,zhang2013kramers}. 
One particularly intriguing class is the two-dimensional chiral topological superconductor (CTSC)~\cite{Read2000,qi2010chiral}. CTSCs are 
renowned for hosting gapless chiral Majorana modes at their boundaries and Majorana zero modes at vortex cores, 
making them a representative platform for topological quantum computation based on Majorana braiding~\cite{Ivanov2001,nayak2008review}.
 
The combination of Rashba SOC (Weyl and Dresselhaus SOC are equivalent in this context, so we focus on Rashba), a Zeeman field, 
and conventional $s$-wave superconductivity provides a well-established route to CTSCs using readily accessible materials~\cite{Sau:2009cqc}. 
Since our bilayer effectively realizes Rashba SOC, CTSCs should emerge when the bilayer is proximitized to an $s$-wave superconductor 
and subjected to a magnetic field. To verify this, we construct the Bogoliubov–de Gennes (BdG) Hamiltonian and map out 
the topological phase diagram. 

\begin{figure}[t]
	\centering
	\includegraphics[width=0.45\textwidth]{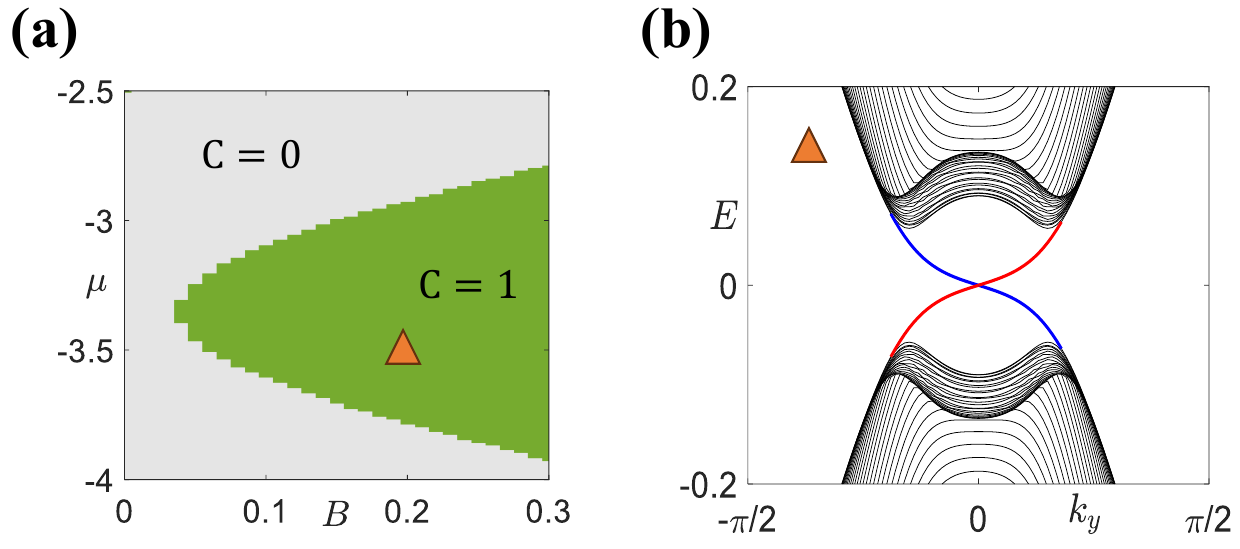}
	\caption{(a) Topological phase diagram of $\mathcal{H}_{\text{BdG}}(\bk)$. (b) Energy spectrum of $\mathcal{H}_{\text{BdG}}(\bk)$ with open (periodic) boundary conditions along the $x$ ($y$) direction, corresponding to the orange triangle at $(\mu,B)=(-3.5,0.2)$ in (a). 
The red (blue) solid line denotes a branch of chiral Majorana edge 
state on the right (left) boundary. Other parameters are: $t_1=1$, $t_2=0.6$, $\lambda=0.25$, 
$M=1$, $\eta=0.5$ and $\Delta=0.55$. }\label{Fig3}
\end{figure}

In the Nambu basis, $\Psi_{\bk}=(\psi_{\bk}^{T},\psi_{-\bk}^{\dag})$
with $\psi_{\bk}=(c_{t,\bk\uparrow},c_{t,\bk\downarrow},c_{b,\bk\uparrow},c_{b,\bk\uparrow})$,
the BdG Hamiltonian is $H=\frac{1}{2}\sum_{\bk}\Psi_{\bk}^{\dag}\mathcal{H}_{\rm BdG}(\bk)\Psi_{\bk}$,
where 
\begin{equation}
	\mathcal{H}_{\rm BdG}(\bk)=\begin{pmatrix}
\mathcal{H}_{\rm R}(\bk)-\mu+Bs_z & i\Delta s_y\\ -i\Delta s_y & -\mathcal{H}_{\rm R}^{T}(-\bk)+\mu-Bs_z
	\end{pmatrix},\label{BDGH}
\end{equation}
Here, $\mu$ is the chemical potential, $B$ the Zeeman field along $z$, and $\Delta$ the proximity-induced $s$-wave pairing gap. 
Numerical diagonalization reveals a fully gapped spectrum except at phase boundaries between 
topologically distinct regimes. Figure \ref{Fig3}(a) shows the topological phase diagram as a function of
$\mu$ and $B$, and Fig.~\ref{Fig3}(b) confirms the existence of gapless chiral Majorana edge states in the topological regime.
The existence of a CTSC with Chern number $C=1$ across a sizable parameter region  demonstrates that our proposed bilayer 
odd-parity magnets operates analogously to conventional SOC systems in realizing CTSCs.
This behavior stands in sharp contrast to odd-parity magnets with unidirectional spin polarization, 
which in two dimensions can only produce topological nodal phases when combined with 
$s$-wave superconductivity~\cite{Nagae2025OPM,Amartya2026,Kim2026OPM,Luo2026OPM}.
Moreover, our setup offers a decisive practical advantage. In conventional SOC systems, 
the topological criterion  $B>\sqrt{(\mu-E_{\rm TRIM})^2+\Delta^2}$ ($E_{\rm TRIM}$ is
the normal-state energy at a time-reversal invariant momentum)
requires the Zeeman field to exceed the pairing gap~\cite{Sato2009TSC,Sau:2009cqc}---a 
stringent constraint because strong fields typically suppress superconductivity. 
As shown in Fig.~\ref{Fig3}(a), our system achieves the topological phase even when  
$B$ is several times smaller than $\Delta$. Thus, the bilayer framework enables 
CTSC realization under substantially more moderate conditions.

{\it Discussions and conclusion.---}We have demonstrated that the nonrelativistic Rashba, 
Weyl, and Dresselhaus SOC realized in our proposed odd-parity coplanar magnets 
bear a striking resemblance to their relativistic counterparts. This resemblance extends 
beyond fundamental properties such as band structure and spin textures to key physical 
consequences, including charge-spin interconversion and the realization of fully-gapped 
topological superconducting phases.
This nonrelativistic mechanism not only provides a route to achieving strong SOC effects 
in light-element materials but also accomplishes what relativistic 
SOC cannot: it enables switching between distinct 
types of SOC---Rashba, Weyl, or Dresselhaus---simply by manipulating the layer N\'{e}el vector.

Our proposed bilayer setup is well within experimental accessibility. The requisite sublattice current order can 
be interaction-driven~\cite{Sun2009loop} or Floquet engineered. In particular, 
several recent studies have shown that circularly polarized light can dynamically induce a 
transition from $\mathcal{PT}$ antiferromagnets to OPAMs~\cite{huang2025oddparityAM,li2025floquet,zhu2025floquet,
liu2025floquet}. The proposed bilayer setup 
can be constructed by following symmetry-guided stacking and using 
materials that have already been predicted~\cite{huang2025oddparityAM,li2025floquet,zhu2025floquet,Pan2025Floquet,Li2025FloquetAM,Tian2026Floquet,Zou2026Floquet,Yu2026AM}. 
A point worth emphasizing is that the circularly 
polarized light driving does not preserve effective time-reversal symmetry, and consequently 
renders the out-of-plane component $\langle s_z(\bk)\rangle$ nonzero. Nevertheless, 
the remaining $[C_{2z}\Vert P]$ symmetry alone is sufficient to protect the in-plane topological 
odd-parity spin textures and the associated key physical consequences.

In conclusion, our study identifies a new class of coplanar magnets whose odd-parity 
spin-polarization components lie in the moment plane and form diverse topologically winding spin textures---a decisive departure from known 
odd-parity coplanar magnets, where the odd-parity component is unidirectionally locked perpendicular to the moment plane. 
Our findings expand the taxonomy of odd-parity coplanar magnets and reveal 
the remarkable breadth of spin-orbit physics accessible through this material class.



{\it Acknowledgements.---}
This work is supported by Fundamental and Interdisciplinary Disciplines 
Breakthrough Plan of the Ministry of Education of China (JYB2025XDXM403), 
Guangdong Basic and Applied Basic Research Foundation (Grant No. 2023B1515040023), 
Natural Science Foundation of China (Grant No. 12474264),
Guangdong Provincial Quantum Science Strategic Initiative (Grant No. GDZX2404007), 
and National Key R\&D Program of China (Grant No. 2022YFA1404103).

\bibliography{dirac.bib}

\begin{widetext}
\clearpage
\begin{center}
\textbf{\large Supplemental Material for ``Nonrelativistic Spin-Orbit-Coupling Effects in Odd-Parity Coplanar Magnets''}\\
\end{center}

\setcounter{equation}{0}
\setcounter{figure}{0}
\setcounter{table}{0}
\makeatletter
\renewcommand{\theequation}{S\arabic{equation}}
\renewcommand{\thefigure}{S\arabic{figure}}
\renewcommand{\bibnumfmt}[1]{[S#1]}

This Supplemental Material comprises three sections, each demonstrating how to analytically determine the spin 
texture of the bilayer Hamiltonian for one specific type. The sections are organized as follows: (I) Rashba 
spin texture, (II) Weyl spin texture, and (III) Dresselhaus spin texture.

\section{I. Rashba spin texture}

We present an analytical approach to derive the spin texture, focusing first on the Rashba case. The full Hamiltonian is given by
\begin{eqnarray}
\mathcal{H}_{R}(\bk)=\left(
              \begin{array}{cc}
                \mathcal{H}_{R}^t(\bk)  & \eta \\
                \eta & \mathcal{H}_{R}^b(\bk) \\
              \end{array}
            \right),\label{bilayerRH}
\end{eqnarray}
where the bottom-layer Hamiltonian reads
\begin{eqnarray}
\mathcal{H}_{R}^b(\bk)=2(t_1\cos k_x+t_2\cos k_y)\sigma_x+4\lambda\sin k_x\cos k_y\sigma_z-Ms_{y}\sigma_z,
\end{eqnarray}
and the top layer Hamiltonian is related to the bottom layer by a $[C_{4z}\Vert C_{4z}]$ symmetry, i.e., 
\begin{eqnarray}
\mathcal{H}_{R}^t(\bk)&=&e^{i\frac{\pi}{4}s_{z}}\mathcal{H}_{R}^b(-k_{y},k_{x})e^{-i\frac{\pi}{4}s_{z}}\nonumber\\
&&=2(t_1\cos k_y+t_2\cos k_x)\sigma_x-4\lambda\sin k_y\cos k_x\sigma_z-Ms_{x}\sigma_z.
\end{eqnarray}
The Pauli matrices $s_{i}$ and $\sigma_{i}$ act on spin and sublattice degrees of freedom, respectively. 
For notational simplicity, identity matrices in both spin and orbital spaces are implicitly assumed unless 
explicitly stated otherwise. Throughout this work, the lattice constants are set to unity.

A well-known hallmark of the Rashba spin texture is that the 
spin polarization direction is perpendicular 
to the momentum vector and exhibits nontrivial winding around a time-reversal invariant momentum. To demonstrate 
that the Bloch bands of $\mathcal{H}_{R}(\bk)$ possess this Rashba texture, we derive the spin texture of 
the lowest-energy band near the Brillouin-zone center, the $\boldsymbol{\Gamma}=(0,0)$ point.

To obtain a low-energy continuum Hamiltonian that captures the spin texture around  $\boldsymbol{\Gamma}$, 
we combine a $\bk \cdot \bp$ expansion with a projection onto the reduced Hilbert space spanned by the 
two eigenstates of the lowest-energy bands at $\boldsymbol{\Gamma}$. We begin with the decoupled limit ($\eta=0$). 
For $\mathcal{H}_{R}^b(\bk)$, the two lowest (negative-energy) bands are degenerate at $\boldsymbol{\Gamma}$
and take the value $E_{-}^{b}=-I_{0}$ with $I_{0}=\sqrt{4(t_{1}+t_{2})^{2}+M^{2}}$. The two corresponding eigenstates at
$\boldsymbol{\Gamma}$ are
\begin{eqnarray}
|u_{R,1}^{b}\rangle&=&|b\rangle\otimes|s_{y}=1\rangle\otimes|\chi_{R,1}\rangle,\nonumber\\
|u_{R,2}^{b}\rangle&=&|b\rangle\otimes|s_{y}=-1\rangle\otimes|\chi_{R,2}\rangle,
\end{eqnarray}
where $|b\rangle$ denotes full polarization to the bottom layer, 
$|s_{y}=\pm1\rangle$ are the two eigenstates of the Pauli matrix $s_{y}$, and 
the spinors $|\chi_{R,1}\rangle$, $|\chi_{R,2}\rangle$ characterize the pseudo-spin polarization of the two sublattices: 
\begin{eqnarray}
|\chi_{R,1}\rangle=\frac{1}{\sqrt{2I_{0}(I_{0}-M)}}\left(
                                                \begin{array}{c}
                                                  2(t_{1}+t_{2}) \\
                                                  -(I_{0}-M) \\
                                                \end{array}
                                              \right),\quad
|\chi_{R,2}\rangle=\frac{1}{\sqrt{2I_{0}(I_{0}+M)}}\left(
                                                \begin{array}{c}
                                                  2(t_{1}+t_{2}) \\
                                                  -(I_{0}+M) \\
                                                \end{array}
                                              \right).
\end{eqnarray}
Similarly, for the top layer, the two lowest-energy bands are also degenerate at $\boldsymbol{\Gamma}$
with the same value $E_{-}^{t}=-I_{0}$. Their eigenstates at $\boldsymbol{\Gamma}$ are 
\begin{eqnarray}
|u_{R,1}^{t}\rangle&=&|t\rangle\otimes|s_{x}=1\rangle\otimes|\chi_{R,1}\rangle,\nonumber\\
|u_{R,2}^{t}\rangle&=&|t\rangle\otimes|s_{x}=-1\rangle\otimes|\chi_{R,2}\rangle.
\end{eqnarray}
Since these two pairs of eigenstates are associated with degenerate eigenenergies, any linear combinations remain eigenstates. 
For the bottom layer, we choose
\begin{eqnarray}
|u_{R,+}^{b}\rangle&=&\frac{1}{\sqrt{2}}(|u_{R,1}^{b}\rangle+|u_{R,2}^{b}\rangle),\nonumber\\
|u_{R,-}^{b}\rangle&=&\frac{1}{\sqrt{2}}(|u_{R,1}^{b}\rangle-|u_{R,2}^{b}\rangle).
\end{eqnarray}
Similarly, for the top layer,
\begin{eqnarray}
|u_{R,+}^{t}\rangle&=&\frac{1}{\sqrt{2}}(|u_{R,1}^{t}\rangle+|u_{R,2}^{t}\rangle),\nonumber\\
|u_{R,-}^{t}\rangle&=&\frac{1}{\sqrt{2}}(|u_{R,1}^{t}\rangle-|u_{R,2}^{t}\rangle).
\end{eqnarray}
These eigenstates have the following property: 
\begin{eqnarray}
\langle u_{R,+}^{\alpha}|s_{z}|u_{R,+}^{\alpha}\rangle&=&\frac{1}{2}(\langle \chi_{R,1}|\chi_{R,2}\rangle+\langle \chi_{R,2}|\chi_{R,1}\rangle)=\frac{\sqrt{I_{0}^{2}-M^{2}}}{I_{0}}\simeq1-\frac{M^{2}}{2I_{0}^{2}}\simeq1,\nonumber\\
\langle u_{R,-}^{\alpha}|s_{z}|u_{R,-}^{\alpha}\rangle&=&-\frac{1}{2}(\langle \chi_{R,1}|\chi_{R,2}\rangle+\langle \chi_{R,2}|\chi_{R,1}\rangle)=-\frac{\sqrt{I_{0}^{2}-M^{2}}}{I_{0}}\simeq-(1-\frac{M^{2}}{2I_{0}^{2}})\simeq-1,\nonumber\\
\langle u_{R,+}^{b}|s_{y}|u_{R,+}^{b}\rangle&=&\langle u_{R,+}^{b}|u_{R,-}^{b}\rangle=0,\nonumber\\
\langle u_{R,-}^{b}|s_{y}|u_{R,-}^{b}\rangle&=&\langle u_{R,-}^{b}|u_{R,+}^{b}\rangle=0,\nonumber\\
\langle u_{R,+}^{b}|s_{x}|u_{R,+}^{b}\rangle&=&\frac{1}{2}(-i\langle \chi_{R,1}|\chi_{R,2}\rangle+i\langle \chi_{R,2}|\chi_{R,1}\rangle)=0,\nonumber\\
\langle u_{R,-}^{b}|s_{x}|u_{R,-}^{b}\rangle&=&\frac{1}{2}(i\langle \chi_{R,1}|\chi_{R,2}\rangle-i\langle \chi_{R,2}|\chi_{R,1}\rangle)=0,\nonumber\\
\langle u_{R,+}^{t}|s_{x}|u_{R,+}^{t}\rangle&=&\langle u_{R,+}^{t}|u_{R,-}^{t}\rangle=0,\nonumber\\
\langle u_{R,-}^{t}|s_{x}|u_{R,-}^{t}\rangle&=&\langle u_{R,-}^{t}|u_{R,+}^{t}\rangle=0,\nonumber\\
\langle u_{R,+}^{t}|s_{y}|u_{R,+}^{t}\rangle&=&\frac{1}{2}(i\langle \chi_{R,1}|\chi_{R,2}\rangle-i\langle \chi_{R,2}|\chi_{R,1}\rangle)=0,\nonumber\\
\langle u_{R,-}^{t}|s_{y}|u_{R,-}^{t}\rangle&=&\frac{1}{2}(-i\langle \chi_{R,1}|\chi_{R,2}\rangle+i\langle \chi_{R,2}|\chi_{R,1}\rangle)=0.
\end{eqnarray}
These results suggest that  $|u_{R,\pm}^{t/b}\rangle$ can be regarded as a spin-up and spin-down basis, respectively. 
In the following, to simplify the derivation we assume $I_{0}\gg M$ and work to linear order in $M/I_{0}$,  
neglecting terms of $O(M^{2}/I_{0}^{2})$.

Near $\boldsymbol{\Gamma}$, the sublattice current order contributes two linear momentum terms, which can be expressed 
as 
\begin{eqnarray}
\mathcal{H}_{R,ssc}(\bk)=-4\lambda k_{y}|t\rangle\langle t|\otimes s_{0}\otimes \sigma_{z}+4\lambda k_{x}|b\rangle\langle b|\otimes s_{0}\otimes \sigma_{z}.
\end{eqnarray}
Here, $s_{0}$ represents the identity matrix in spin space. Projecting it onto the basis $(|u_{R,+}^{t}\rangle,|u_{R,-}^{t}\rangle,|u_{R,+}^{b}\rangle,|u_{R,-}^{b}\rangle)$, we obtain
\begin{eqnarray}
\widetilde{\mathcal{H}}_{R,ssc}(\bk)=\frac{4\lambda M}{I_{0}}\left(
                                                             \begin{array}{cccc}
                                                               0 & -k_{y} & 0 & 0 \\
                                                               -k_{y} & 0 & 0 & 0 \\
                                                               0 & 0 & 0 & k_{x} \\
                                                               0 & 0 &  k_{x} & 0 \\
                                                             \end{array}
                                                           \right).
\end{eqnarray}
On the other hand, the interlayer coupling term can be expressed as 
\begin{eqnarray}
\mathcal{H}_{R,lc}(\bk)=\eta(|t\rangle\langle b|+|b\rangle\langle t|)\otimes s_{0}\otimes \sigma_{z}\equiv\eta\tau_{x}\otimes s_{0}\otimes \sigma_{z},
\end{eqnarray}
where $\tau_{i}$ are Pauli matrices acting on the two layer degrees of freedom. Projecting onto the same basis as above, we obtain
\begin{eqnarray}
\widetilde{\mathcal{H}}_{R,lc}(\bk)=\eta\left(
                                                             \begin{array}{cccc}
                                                               0 & 0 & 1 & 0 \\
                                                               0 & 0 & 0 & i \\
                                                               1 & 0 & 0 & 0 \\
                                                               0 & -i & 0 & 0 \\
                                                             \end{array}
                                                           \right).
\end{eqnarray}
Consequently, we obtain a four-band effective Hamiltonian that describes the four negative bands near $\boldsymbol{\Gamma}$:
\begin{eqnarray}
\mathcal{H}_{\rm R, eff}(\bk)=-I_{0}-\frac{4\lambda M}{I_{0}} k_{y}\frac{\rho_{z}+1}{2}s_{x}+\frac{4\lambda M}{I_{0}} k_{x}\frac{1-\rho_{z}}{2}s_{x}
+\frac{\eta}{2}(\rho_{x}-\rho_{y})+\frac{\eta}{2}(\rho_{x}+\rho_{y})s_{z},\label{effectiveRH}
\end{eqnarray}
where the Pauli matrix $s_{i}$ still acts on the spin space, while the Pauli matrices $\rho_{i}$ 
act on a layer-sublattice mixing subspace. 

To further derive the low-energy Hamiltonian describing the lowest-energy pair of bands, we follow similar procedures as above. At $\boldsymbol{\Gamma}$, the Hamiltonian $\mathcal{H}_{\rm R, eff}$ possesses two pairs of degenerate eigenvalues: one at 
$E_{1}=-I_{0}-\eta$  and the other at $E_{2}=-I_{0}+\eta$. The two eigenstates corresponding to $E_{1}$ are 
\begin{eqnarray}
|v_{R,1}\rangle=|\xi_{R,1}\rangle\otimes|s_{z}=1\rangle, \quad
|v_{R,2}\rangle=|\xi_{R,2}\rangle\otimes|s_{z}=-1\rangle,
\end{eqnarray}
where 
\begin{eqnarray}
|\xi_{R,1}\rangle=\frac{1}{\sqrt{2}}\left(\begin{array}{c}
                1 \\
                -1 
              \end{array}\right),\quad
|\xi_{R,2}\rangle=\frac{1}{\sqrt{2}}\left(\begin{array}{c}
                1 \\
                i
              \end{array}\right).           
\end{eqnarray}
Projecting the two linear momentum terms in Hamiltonian (\ref{effectiveRH}) onto the two-dimensional Hilbert space
spanned by $|v_{R,1}\rangle$ and  $|v_{R,2}\rangle$, we find 
\begin{eqnarray}
\langle v_{R,m}|-\frac{4\lambda M}{I_{0}} k_{y}\frac{\rho_{z}+1}{2}s_{x}|v_{R,n}\rangle&=&-\frac{4\lambda M}{I_{0}} k_{y}\left(
                                                                                 \begin{array}{cc}
                                                                                   0 & \langle \xi_{R,1}|\frac{\rho_{z}+1}{2}|\xi_{R,2}\rangle \\
                                                                                   \langle \xi_{R,2}|\frac{\rho_{z}+1}{2}|\xi_{R,1}\rangle & 0 \\
                                                                                 \end{array}
                                                                                 \right)_{mn}=-(\frac{2\lambda M}{I_{0}}k_{y}s_{x})_{mn}\nonumber\\
 \langle v_{R,m}|\frac{4\lambda M}{I_{0}} k_{x}\frac{1-\rho_{z}}{2}s_{x}|v_{R,n}\rangle&=&\frac{4\lambda M}{I_{0}} k_{x}\left(
                                                                                 \begin{array}{cc}
                                                                                   0 & \langle \xi_{R,1}|\frac{1-\rho_{z}}{2}|\xi_{R,2}\rangle \\
                                                                                   \langle \xi_{R,2}|\frac{1-\rho_{z}}{2}|\xi_{R,1}\rangle & 0 \\
                                                                                 \end{array}
                                                                               \right)_{mn}=(\frac{2\lambda M}{I_{0}}k_{x}s_{y})_{mn}                                                                            
\end{eqnarray}
Therefore, to linear order in momentum,  the lowest-energy pair (LEP) of  bands near $\boldsymbol{\Gamma}$ can be effectively described by
\begin{eqnarray}
\mathcal{H}_{\rm R,LEP}(\bk)=-I_{0}-\eta+\frac{2\lambda M}{I_{0}}(k_{x}s_{y}-k_{y}s_{x}). \label{lowenergyRH}
\end{eqnarray} 
While the term $(k_{x}s_{y}-k_{y}s_{x})$ takes the standard form of Rashba spin-orbit coupling (SOC), this does not 
necessarily imply that the underlying spin texture is of Rashba type. The reason is that the basis functions 
are not simply $|s_{z}=1\rangle$
and $|s_{z}=-1\rangle$, but also involve layer and sublattice degrees of freedom.  Moreover, the explicit 
form of $\mathcal{H}_{\rm R, LEP}(\bk)$ depends on the choice of basis for the projection. If we perform a 
gauge transformation on the two basis functions $|v_{R,1}\rangle$ and  $|v_{R,2}\rangle$, we will generally 
obtain a different expression for $\mathcal{H}_{\rm R, LEP}(\bk)$. For instance, if we choose the basis as 
$(|\widetilde{v}_{R,1}\rangle, |\widetilde{v}_{R,2}\rangle)=(|v_{R,1}\rangle,i|v_{R,2}\rangle)$, we obtain 
\begin{eqnarray}
\widetilde{\mathcal{H}}_{\rm R, LEP}(\bk)=-I_{0}-\eta+\frac{2\lambda M}{I_{0}}(k_{x}s_{x}+k_{y}s_{y}). 
\end{eqnarray} 
Clearly, the SOC term now takes a Weyl-type form. Therefore, to determine the actual 
type of SOC realized, we must compute the spin texture, which is independent of gauge choice.

To calculate the spin texture, we need to retain the wavefunction information at each step of the projection. 
Let us focus on the spin texture of the lowest-energy band. From the Hamiltonian (\ref{lowenergyRH}), the 
eigenstate of the lower band is
\begin{eqnarray}
|\psi_{R,-}(\bk)\rangle=\frac{1}{\sqrt{2}}\left(\begin{array}{c}
                1 \\
                -ie^{i\theta_{\bk}} 
              \end{array}\right),
\end{eqnarray}
where $\theta_{\bk}=\arg(k_{x}+ik_{y})$. Taking into account the basis functions at each projection step,
 this eigenstate in the original eight-band basis can be expressed as
\begin{eqnarray}
|\psi_{R,-}(\bk)\rangle&=&\frac{1}{\sqrt{2}}[|\xi_{R,1}\rangle\otimes|s_{z}=1\rangle-ie^{i\theta_{\bk}}|\xi_{R,2}\rangle\otimes|s_{z}=-1\rangle]\nonumber\\
&=&\frac{1}{2}[(|u_{R,+}^{t}\rangle-|u_{R,+}^{b}\rangle)-ie^{i\theta_{\bk}}(|u_{R,-}^{t}\rangle+i|u_{R,-}^{b}\rangle)].
\end{eqnarray}
Consequently, the spin texture is given by
\begin{eqnarray}
\langle s_{x}(\bk)\rangle&\equiv&\langle \psi_{R,-}(\bk)|s_{x}|\psi_{R,-}(\bk)\rangle\nonumber\\
&=&\frac{1}{4}[-ie^{i\theta_{\bk}}\langle u_{R,+}^{t}|s_{x}|u_{R,-}^{t}\rangle+ie^{-i\theta_{\bk}}\langle u_{R,-}^{t}|s_{x}|u_{R,+}^{t}\rangle
-e^{i\theta_{\bk}}\langle u_{R,+}^{b}|s_{x}|u_{R,-}^{b}\rangle-e^{-i\theta_{\bk}}\langle u_{R,-}^{b}|s_{x}|u_{R,+}^{b}\rangle]\nonumber\\
&=&\frac{1}{4}[-ie^{i\theta_{\bk}}+ie^{-i\theta_{\bk}}-\frac{1}{2}ie^{i\theta_{\bk}}(\langle \chi_{R,1}|\chi_{R,2}\rangle+\langle \chi_{R,2}|\chi_{R,1}\rangle)+\frac{1}{2}ie^{-i\theta_{\bk}}(\langle \chi_{R,1}|\chi_{R,2}\rangle+\langle \chi_{R,2}|\chi_{R,1}\rangle)]\nonumber\\
&=&\frac{1}{2}[1+\frac{\sqrt{I_{0}^{2}-M^{2}}}{I_{0}}]\sin\theta_{\bk}\nonumber\\
&\simeq&\sin\theta_{\bk},\nonumber\\
\langle s_{y}(\bk)\rangle&\equiv&\langle \psi_{R,-}(\bk)|s_{y}|\psi_{R,-}(\bk)\rangle\nonumber\\
&=&\frac{1}{4}[-ie^{i\theta_{\bk}}\langle u_{R,+}^{t}|s_{y}|u_{-}^{t}\rangle+ie^{-i\theta_{\bk}}\langle u_{R,-}^{t}|s_{y}|u_{R,+}^{t}\rangle
-e^{i\theta_{\bk}}\langle u_{R,+}^{b}|s_{y}|u_{R,-}^{b}\rangle-e^{-i\theta_{\bk}}\langle u_{R,-}^{b}|s_{y}|u_{R,+}^{b}\rangle]\nonumber\\
&=&\frac{1}{4}[-\frac{1}{2}e^{i\theta_{\bk}}(\langle \chi_{R,1}|\chi_{R,2}\rangle+\langle \chi_{R,2}|\chi_{R,1}\rangle)-\frac{1}{2}e^{-i\theta_{\bk}}(\langle \chi_{R,1}|\chi_{R,2}\rangle+\langle \chi_{R,2}|\chi_{R,1}\rangle)-e^{i\theta_{\bk}}-e^{-i\theta_{\bk}}]\nonumber\\
&=&-\frac{1}{2}[1+\frac{\sqrt{I_{0}^{2}-M^{2}}}{I_{0}}]\cos\theta_{\bk}\nonumber\\
&\simeq&-\cos\theta_{\bk},\nonumber\\
\langle s_{z}(\bk)\rangle&\equiv&\langle \psi_{R,-}(\bk)|s_{z}|\psi_{R,-}(\bk)\rangle=0.
\end{eqnarray}
Evidently, the in-plane spin texture $(\langle s_{x}(\bk)\rangle, \langle s_{y}(\bk)\rangle) = (\sin\theta_{\bk}, -\cos\theta_{\bk})$ is always perpendicular to the momentum vector $\bk = k(\cos\theta_{\bk}, \sin\theta_{\bk})$. This is precisely the hallmark of a 
Rashba spin texture, demonstrating that the bilayer  Hamiltonian (\ref{bilayerRH}) produces a Rashba spin texture in its energy bands.

\section{II. Weyl spin texture}

The Hamiltonian that gives rise to a Weyl spin texture reads
\begin{eqnarray}
\mathcal{H}_{W}(\bk)=\left(
              \begin{array}{cc}
                \mathcal{H}_{W}^t(\bk)  & \eta \\
                \eta & \mathcal{H}_{W}^b(\bk) \\
              \end{array}
            \right),
\end{eqnarray}
where the bottom-layer Hamiltonian is
\begin{eqnarray}
\mathcal{H}_{W}^b(\bk)=2(t_1\cos k_x+t_2\cos k_y)\sigma_x+4\lambda\sin k_x\cos k_y\sigma_z+Ms_{x}\sigma_z,
\end{eqnarray}
and the top layer Hamiltonian is related to the bottom layer by a $[C_{4z}\Vert C_{4z}]$ symmetry, i.e, 
\begin{eqnarray}
\mathcal{H}_{W}^t(\bk)&=&e^{i\frac{\pi}{4}s_{z}}\mathcal{H}^b(-k_{y},k_{x})e^{-i\frac{\pi}{4}s_{z}}\nonumber\\
&&=2(t_1\cos k_y+t_2\cos k_x)\sigma_x-4\lambda\sin k_y\cos k_x\sigma_z-Ms_{y}\sigma_z.
\end{eqnarray}
Compared to the Rashba case, the Hamiltonian in each layer differs only by a  $90^{\circ}$
anticlockwise rotation of the N\'{e}el vector: specifically, bottom:  $-M\hat{y}\rightarrow M\hat{x}$, 
top: $-M\hat{x}\rightarrow -M\hat{y}$. 

Following the same procedure as before, we first write down the eigenstates of the four lowest-energy bands at 
$\boldsymbol{\Gamma}$:
\begin{eqnarray}
|u_{W,+}^{t}\rangle&=&\frac{1}{\sqrt{2}}(|u_{W,1}^{t}\rangle+|u_{W,2}^{t}\rangle),\quad
|u_{W,-}^{t}\rangle=\frac{1}{\sqrt{2}}(|u_{W,1}^{t}\rangle-|u_{W,2}^{t}\rangle),\nonumber\\
|u_{W,+}^{b}\rangle&=&\frac{1}{\sqrt{2}}(|u_{W,1}^{b}\rangle+|u_{W,2}^{b}\rangle),\quad
|u_{W,-}^{b}\rangle=\frac{1}{\sqrt{2}}(|u_{W,1}^{b}\rangle-|u_{W,2}^{b}\rangle),
\end{eqnarray}
where
\begin{eqnarray}
|u_{W,1}^{t}\rangle&=&|t\rangle\otimes|s_{y}=1\rangle\otimes|\chi_{W,2}\rangle,\quad
|u_{W,2}^{t}\rangle=|t\rangle\otimes|s_{y}=-1\rangle\otimes|\chi_{W,1}\rangle,\nonumber\\
|u_{W,1}^{b}\rangle&=&|b\rangle\otimes|s_{x}=1\rangle\otimes|\chi_{W,1}\rangle,\quad
|u_{W,2}^{b}\rangle=|b\rangle\otimes|s_{x}=-1\rangle\otimes|\chi_{W,2}\rangle,
\end{eqnarray}
with
\begin{eqnarray}
|\chi_{W,1}\rangle&=&\frac{1}{\sqrt{2I_{0}(I_{0}+M)}}\left(
                                                \begin{array}{c}
                                                  2(t_{1}+t_{2}) \\
                                                  -(I_{0}+M) \\
                                                \end{array}
                                              \right)=|\chi_{R,2}\rangle,\nonumber\\
|\chi_{W,2}\rangle&=&\frac{1}{\sqrt{2I_{0}(I_{0}-M)}}\left(
                                                \begin{array}{c}
                                                  2(t_{1}+t_{2}) \\
                                                  -(I_{0}-M) \\
                                                \end{array}
                                              \right)=|\chi_{R,1}\rangle.
\end{eqnarray}

Projecting the sublattice current term and the interlayer coupling term  onto the basis 
$(|u_{W,+}^{t}\rangle,|u_{W,-}^{t}\rangle,|u_{W,+}^{b}\rangle,|u_{W,-}^{b}\rangle)$,
we obtain 
\begin{eqnarray}
\widetilde{\mathcal{H}}_{W,ssc}(\bk)=\frac{4\lambda M}{I_{0}}\left(
                                                             \begin{array}{cccc}
                                                               0 & - k_{y} & 0 & 0 \\
                                                               - k_{y} & 0 & 0 & 0 \\
                                                               0 & 0 & 0 & - k_{x} \\
                                                               0 & 0 & - k_{x} & 0 \\
                                                             \end{array}
                                                           \right),
\end{eqnarray}
and 
\begin{eqnarray}
\widetilde{\mathcal{H}}_{W,lc}(\bk)=\eta\left(
                                                             \begin{array}{cccc}
                                                               0 & 0 & 1 & 0 \\
                                                               0 & 0 & 0 & -i \\
                                                               1 & 0 & 0 & 0 \\
                                                               0 & i & 0 & 0 \\
                                                             \end{array}
                                                           \right).
\end{eqnarray}

Accordingly, the effective Hamiltonian describing the four lowest-energy bands reads 
\begin{eqnarray}
\mathcal{H}_{\rm W, eff}(\bk)=-I_{0}-\frac{4\lambda M}{I_{0}} k_{y}\frac{\rho_{z}+1}{2}s_{x}-\frac{4\lambda M}{I_{0}} k_{x}\frac{1-\rho_{z}}{2}s_{x}
+\frac{\eta}{2}(\rho_{x}+\rho_{y})+\frac{\eta}{2}(\rho_{x}-\rho_{y})s_{z}.
\end{eqnarray} 
At $\boldsymbol{\Gamma}$, $\mathcal{H}_{\rm W, eff}(\bk)$ has two pairs of degenerate eigenvalues, one at 
$E_{1}=-I_{0}-\eta$  and the other $E_{2}=-I_{0}+\eta$.
The two eigenstates corresponding to $E_{1}$ are 
\begin{eqnarray}
|v_{W,1}\rangle=|\xi_{W,1}\rangle\otimes|s_{z}=1\rangle, \quad
|v_{W,2}\rangle=|\xi_{W,2}\rangle\otimes|s_{z}=-1\rangle,
\end{eqnarray}
where 
\begin{eqnarray}
|\xi_{W,1}\rangle=\frac{1}{\sqrt{2}}\left(\begin{array}{c}
                1 \\
                -1 
              \end{array}\right),\quad
|\xi_{W,2}\rangle=\frac{1}{\sqrt{2}}\left(\begin{array}{c}
                1 \\
                -i
              \end{array}\right).           
\end{eqnarray}
Projecting the two linear momentum terms onto the basis spanned by $|v_{W,1}\rangle$ and $|v_{W,2}\rangle$, 
we obtain
\begin{eqnarray}
\langle v_{W,m}|-\frac{4\lambda M}{I_{0}} k_{y}\frac{\rho_{z}+1}{2}s_{x}|v_{W,n}\rangle&=&-\frac{4\lambda M}{I_{0}} k_{y}\left(
                                                                                 \begin{array}{cc}
                                                                                   0 & \langle \xi_{W,1}|\frac{\rho_{z}+1}{2}|\xi_{W,2}\rangle \\
                                                                                   \langle \xi_{W,2}|\frac{\rho_{z}+1}{2}|\xi_{W,1}\rangle & 0 \\
                                                                                 \end{array}
                                                                                 \right)_{mn}=(-\frac{2\lambda M}{I_{0}}k_{y}s_{x})_{mn},\nonumber\\
 \langle v_{W,m}|-\frac{4\lambda M}{I_{0}} k_{x}\frac{1-\rho_{z}}{2}s_{x}|v_{W,n}\rangle&=&-\frac{4\lambda M}{I_{0}} k_{x}\left(
                                                                                 \begin{array}{cc}
                                                                                   0 & \langle \xi_{W,1}|\frac{1-\rho_{z}}{2}|\xi_{W,2}\rangle \\
                                                                                   \langle \xi_{W,2}|\frac{1-\rho_{z}}{2}|\xi_{W,1}\rangle & 0 \\
                                                                                 \end{array}
                                                                               \right)_{mn}=(\frac{2\lambda M}{I_{0}}k_{x}s_{y})_{mn}.                                                                            
\end{eqnarray}
Therefore, to the linear order in momentum,  the lowest-energy pair of bands near $\boldsymbol{\Gamma}$ can be effectively described by
\begin{eqnarray}
\mathcal{H}_{\rm W, LEP}(\bk)=-I_{0}-\eta+\frac{2\lambda M}{I_{0}}(k_{x}s_{y}-k_{y}s_{x}). \label{lowestWeyl}
\end{eqnarray} 
Although the SOC term takes a form identical to that of Rashba SOC, the resulting spin texture is 
not necessarily of Rashba type, for reasons previously explained. To explicitly demonstrate that 
the spin texture is in fact Weyl-type, we write down the eigenstate of the lower band:
\begin{eqnarray}
|\psi_{W,-}(\bk)\rangle=\frac{1}{\sqrt{2}}\left(\begin{array}{c}
                1 \\
                -ie^{i\theta_{\bk}} 
              \end{array}\right).
\end{eqnarray}
Taking into account the basis functions at each step of the projection, this eigenstate in the original eight-band basis can be expressed as
\begin{eqnarray}
|\psi_{W,-}(\bk)\rangle&=&\frac{1}{\sqrt{2}}[|\xi_{W,1}\rangle\otimes|s_{z}=1\rangle-ie^{i\theta_{\bk}}|\xi_{W,2}\rangle\otimes|s_{z}=-1\rangle]\nonumber\\
&=&\frac{1}{2}[(|u_{W,+}^{t}\rangle-|u_{W,+}^{b}\rangle)-ie^{i\theta_{\bk}}(|u_{W,-}^{t}\rangle-i|u_{W,-}^{b}\rangle)].
\end{eqnarray}
A straightforward calculation then yields
\begin{eqnarray}
\langle s_{x}(\bk)\rangle&\equiv&\langle \psi_{W,-}(\bk)|s_{x}|\psi_{W,-}(\bk)\rangle\nonumber\\
&=&\frac{1}{4}[-ie^{i\theta_{k}}\langle u_{W,+}^{t}|s_{x}|u_{W,-}^{t}\rangle+ie^{-i\theta_{k}}\langle u_{W,-}^{t}|s_{x}|u_{W,+}^{t}\rangle
+e^{i\theta_{k}}\langle u_{W,+}^{b}|s_{x}|u_{W,-}^{b}\rangle+e^{-i\theta_{k}}\langle u_{W,-}^{b}|s_{x}|u_{W,+}^{b}\rangle]\nonumber\\
&=&\frac{1}{4}[\frac{1}{2}e^{i\theta_{k}}(\langle \chi_{W,1}|\chi_{W,2}\rangle+\langle \chi_{W,2}|\chi_{W,1}\rangle)+\frac{1}{2}e^{-i\theta_{k}}(\langle \chi_{W,1}|\chi_{W,2}\rangle+\langle \chi_{W,2}|\chi_{W,1}\rangle)+e^{i\theta_{k}}+e^{-i\theta_{k}}]\nonumber\\
&=&\frac{1}{2}[1+\frac{\sqrt{I_{0}^{2}-M^{2}}}{I_{0}}]\cos\theta_{\bk}\nonumber\\
&\simeq&\cos\theta(\bk),\nonumber\\
\langle s_{y}(\bk)\rangle&\equiv&\langle \psi_{W,-}(\bk)|s_{y}|\psi_{W,-}(\bk)\rangle\nonumber\\
&=&\frac{1}{4}[-ie^{i\theta_{k}}\langle u_{W,+}^{t}|s_{y}|u_{W,-}^{t}\rangle+ie^{-i\theta_{k}}\langle u_{W,-}^{t}|s_{y}|u_{W,+}^{t}\rangle
+e^{i\theta_{k}}\langle u_{W,+}^{b}|s_{y}|u_{W,-}^{b}\rangle+e^{-i\theta_{k}}\langle u_{W,-}^{b}|s_{y}|u_{W,+}^{b}\rangle]\nonumber\\
&=&\frac{1}{4}[-ie^{i\theta_{k}}+ie^{-i\theta_{k}}-\frac{1}{2}ie^{i\theta_{k}}(\langle \chi_{W,1}|\chi_{W,2}\rangle+\langle \chi_{W,2}|\chi_{W,1}\rangle)+\frac{1}{2}ie^{-i\theta_{k}}(\langle\chi_{W,1}|\chi_{W,2}\rangle+\langle \chi_{W,2}|\chi_{W,1}\rangle)]\nonumber\\
&=&\frac{1}{2}[1+\frac{\sqrt{I_{0}^{2}-M^{2}}}{I_{0}}]\sin\theta_{\bk}\nonumber\\
&\simeq&\sin\theta(\bk),\nonumber\\
\langle s_{z}(\bk)\rangle&\equiv&\langle \psi_{W,-}(\bk)|s_{z}|\psi_{W,-}(\bk)\rangle=0.
\end{eqnarray}
Evidently, the in-plane spin texture $(\langle s_{x}(\bk)\rangle, \langle s_{y}(\bk)\rangle) = (\cos\theta_{\bk}, \sin\theta_{\bk})$ 
is always parallel to the momentum vector $\bk = k(\cos\theta_{\bk}, \sin\theta_{\bk})$. This is precisely the hallmark of a 
Weyl spin texture, confirming that the energy bands of this Hamiltonian indeed exhibit Weyl spin texture.

Before concluding this section, we remark on why the two low-energy Hamiltonians [Eq.~(\ref{lowenergyRH}) 
and Eq.~(\ref{lowestWeyl})] describing the lowest-energy
pair of bands for the Rashba and Weyl cases can be identical given the basis we have chosen. The underlying 
reason is that Rashba and Weyl spin textures are topologically equivalent, as they are characterized by the same winding number. 
Specifically, one can perform a unitary rotation in spin space to transform one type of spin texture into the other.

\section{III. Dresselhaus spin texture}

The Hamiltonian that produces a Dresselhaus spin texture reads
\begin{eqnarray}
\mathcal{H}_{D}(\bk)=\left(
              \begin{array}{cc}
                \mathcal{H}_{D}^t(\bk)  & \eta \\
                \eta & \mathcal{H}_{D}^b(\bk) \\
              \end{array}
            \right),
\end{eqnarray}
where the bottom layer Hamiltonian becomes
\begin{eqnarray}
\mathcal{H}_{D}^b(\bk)=2(t_1\cos k_x+t_2\cos k_y)\sigma_x+4\lambda\sin k_x\cos k_y\sigma_z+Ms_{x}\sigma_z,
\end{eqnarray}
and the top layer Hamiltonian becomes 
\begin{eqnarray}
\mathcal{H}_{D}^t(\bk)&=&e^{i\frac{\pi}{4}s_{z}}\mathcal{H}^b(k_{y},-k_{x})e^{-i\frac{\pi}{4}s_{z}}\nonumber\\
&&=2(t_1\cos k_y+t_2\cos k_x)\sigma_x+4\lambda\sin k_y\cos k_x\sigma_z-Ms_{y}\sigma_z.
\end{eqnarray}
Compared to the Weyl case, the only difference is that the term describing the sublattice
current order in the top layer takes the opposite sign. Consequently, the spin texture for this case 
can be obtained directly by replacing $(k_{x},k_{y})$ by $(k_{x},-k_{y})$ at all steps of the derivation. 
Under this substitution, $\theta_{\bk}$ is accordingly replaced by $-\theta_{\bk}$. Therefore, the spin texture 
for the lowest-energy band of this Hamiltonian is
\begin{eqnarray}
\langle s_{x}(\bk)&\simeq&\cos(-\theta(\bk))=\cos(\theta(\bk)),\nonumber\\
\langle s_{y}(\bk)\rangle&\simeq&\sin(-\theta(\bk))=-\sin(\theta(\bk)),\nonumber\\
\langle s_{z}(\bk)\rangle&=&0.
\end{eqnarray}
The spin-texture pattern is identical to that induced by the Dresselhaus SOC proportional 
to $(k_{x}s_{x}-k_{y}s_{y})$, confirming that the energy bands of this Hamiltonian 
indeed possess a Dresselhaus spin texture.

\end{widetext}

\end{document}